\documentclass[12pt]{article}
\pdfoutput=1
\usepackage{dcolumn}
\usepackage{bm}
\usepackage{graphicx}
\usepackage{amssymb,amsmath}
\usepackage{multirow}
\usepackage{cite,color,url}
\usepackage{tabu}
\usepackage{array}
\usepackage[colorlinks=true,urlcolor=blue,anchorcolor=blue
,citecolor=blue,filecolor=blue,linkcolor=blue,menucolor=blue
,linktocpage=true,pdfproducer=medialab,pdfa=true]{hyperref}

\usepackage{colordvi}
\usepackage{epsfig,psfrag,rotating,soul}
\def\beqa{\begin{eqnarray}}
\def\eeqa{\end{eqnarray}}
\newcommand{\be}{\ensuremath{\beta}}
\newcommand{\al}{\ensuremath{\alpha}}

\evensidemargin \oddsidemargin
\allowdisplaybreaks

\psfrag{lk}{$l_k$}
\psfrag{lm}{$l_m$}
\psfrag{blk}{$\bar {l}_k$}
\psfrag{blm}{$\bar {l}_m$}
\def\thefootnote{\fnsymbol{footnote}}

\let\OLDthebibliography\thebibliography
\renewcommand\thebibliography[1]{
  \OLDthebibliography{#1}
  \setlength{\parskip}{0pt}
  \setlength{\itemsep}{0pt plus 0.3ex}}
\begin{document}
\thispagestyle{empty}
\vspace*{2.5cm}
\vspace{0.5cm}
\begin{center}

\begin{Large}
\textbf{\textsc{Triple Higgs coupling effect on $h^0 \to b \bar b$ and 
$h^0 \to \tau^+ \tau^-$ in the 2HDM}}
\end{Large}

\vspace{1cm}
{\sc
A. Arhrib$^{1}$%
\footnote{\tt \href{mailto:aarhrib@gmail.com}{aarhrib@gmail.com}}%
, 
R. Benbrik$^{2}$%
\footnote{\tt
\href{mailto:r.benbrik@uca.ma}{r.benbrik@uca.ac.ma}}
,
J. El Falaki$^{1}$%
\footnote{\tt
\href{mailto:jaouad.elfalaki@gmail.com}{jaouad.elfalaki@gmail.com}},
W. Hollik$^{3}$%
\footnote{\tt
\href{mailto:hollik@mpp.mpg.de}{hollik@mpp.mpg.de}},
}

\vspace*{.7cm}
{\sl
$^1$ Abdelmalek Essaadi University, Faculty of Sciences and Techniques,
Tanger, Morocco}

{\sl
$^2$ MSISM Team, Facult\'e Polydisciplinaire de Safi,
Sidi Bouzid, B.P. 4162,  Safi, Morocco}

{\sl
$^3$ Max Planck Institut f\"ur Physik, F\"ohringer Ring 6, 80805 M\"unchen, Germany}

\end{center}

\vspace*{0.1cm}

\begin{abstract}
\noindent
We study the one-loop electroweak radiative corrections to   
 $h^0\to b\bar{b}$ and $h^0\to \tau^+\tau^-$
in the framework of two Higgs doublet Model (2HDM). We evaluate the deviation
of these couplings from their Standard Model (SM) values. 
$h^0\to b\bar{b}$ and $h^0\to \tau^+\tau^-$ may
receives large contribution 
from triple Higgs couplings $h^0H^0H^0$, $H^0h^0h^0$, $h^0A^0A^0$ and $h^0H^+H^-$ 
which are absent in the Standard Model. 
It is found that in 2HDM, these corrections could be significant 
and may reach more than 12\% for not tow heavy $H^0$ or $A^0$ or $H^\pm$.
We also study the ratio of branching ratios
$R=BR(h^0\to b\bar{b})/BR(h^0\to \tau^+\tau^-)$ of Higgs
boson decays which could be used to  disentangle SM from other 
models such as 2HDM. 
\end{abstract}

\def\thefootnote{\arabic{footnote}}
\setcounter{page}{0}
\setcounter{footnote}{0}
\newpage

\section{Introduction}
\label{intro}

A Higgs-like particle has been discovered in the first run of the LHC 
with 7 and 8 TeV  energy in 2012 \cite{cmsdiscovery,atlasdiscovery}. 
The combined measured Higgs boson mass  
obtained by the ATLAS and CMS  collaborations based on the data from 7 and 8 TeV
is $m_{h} =$ 125.09 $\pm$ 0.21 (stat.) $\pm$ 0.11 (syst.) 
GeV~\cite{Aad:2015zhl}. 
ATLAS and CMS also performed several Higgs coupling measurements, such as
 Higgs couplings to $W^+W^-$, $ZZ$, $\gamma\gamma$, $b\bar{b}$ 
and $\tau^+ \tau^-$ with 20-30\% uncertainty, while the coupling to 
 $b\bar{b}$ still suffers from a large uncertainty of $40-50$\%. 
One of the tasks of the new LHC run at 13 TeV (and 14 TeV) would be to
improve all the aforementioned measurements and to perform new ones
such as accessing
$h^0\to \gamma Z$ as well as the triple self-coupling of the Higgs boson. 
It is expected that the new LHC run  will pin down the 
uncertainty in $h^0 \to b \bar b$ and $h^0 \to \tau^+ \tau^-$ to 
$10$-$13\%$ and $6$-$8\%$ for bottom quarks and tau leptons, respectively. 
These measurements will be further ameliorated by the High Luminosity 
 option for the LHC  (HL-LHC) down to uncertainties of $4$-$7\%$ for 
$b$ quarks and $2$-$5\%$ for  $\tau$ leptons \cite{accuracy1}. 
Moreover, in the clean
environment of the $e^+e^-$ Linear Collider (LC), which can act 
as a Higgs factory,  the uncertainties on $h \to b \bar b$
 and $h^0 \to \tau^+ \tau^-$ would be much smaller reaching  
$0.6\%$ for the couplings in $h^0\to b \bar b$ 
and $1.3\%$ for those in $h^0\to \tau^+ \tau^-$~\cite{accuracy2,accuracy3}.

The above accuracies on fermionic Higgs decay measurements, if reached, 
are of the size comparable to the effects of radiative corrections to some Higgs
decays.  Therefore, one can use these radiative correction effects 
to distinguish between the Standard Model (SM) and various beyond-standard models.
In this respect, precise calculations of Higgs-boson production and decay rates have 
been performed already quite some time ago with great 
achievements (see e.g.~\cite{Djouadi:2005gi,Gunion:1989we}). 
QCD corrections to Higgs decays into quarks 
are very well known up to ${\cal O}(\alpha_s^3)$ 
as well as additional corrections at  ${\cal O}(\alpha_s^2)$ 
that involve logarithms of the light-quark masses
 and also heavy top contributions \cite{Djouadi:2005gi}.
Electroweak radiative corrections to  fermionic 
decays ($b\bar{b}$ and $\tau^+\tau^-$) of the Higgs boson in the 
SM are also well established~\cite{Dabelstein:1991ky,Bardin:1990zj,Kniehl:1991ze} 
in the on-shell scheme. 
In the framework of the Two-Higgs-Doublet Model
(2HDM), several studies have been 
carried out to evaluate the electroweak corrections to fermionic Higgs decays
\cite{Arhrib:2003ph,Kanemura:2014dja}. 
The calculation of Ref.~\cite{Arhrib:2003ph} is done in the on-shell scheme
except for the Higgs field renormalization where the $\overline{MS}$
 subtraction has been used, while the one of Ref.~\cite{Kanemura:2014dja} 
 is performed using the on-shell renormalization scheme of 
\cite{Kanemura:2004mg}. 

In this paper, we will study the  effects of electroweak radiative corrections 
to $h^0 \to b \bar b$ and $h^0 \to \tau^+ \tau^-$ decays in the 2HDM 
taking into account theoretical constraints 
as well as experimental restrictions from recent LHC data and other 
experimental results. 
For $h^0\to b \bar b$ we will update our results from
 \cite{Arhrib:2003ph}  while for $h^0 \to \tau^+ \tau^-$ we will compute 
these effects for the first time following the same renormalization  
procedure described in \cite{Arhrib:2003ph}. 
Similar studies have been performed in
\cite{Kanemura:2015fra, Kanemura:2014bqa} 
 to which we will compare our results. 
 We will also use our calculations to
evaluate the ratio of branching fractions of Higgs decays in the 2HDM 
~\cite{Guasch:2001wv,Arganda:2015bta},
\begin{eqnarray}
R=\frac{BR(h^0 \to b\bar{b})}{BR(h^0 \to \tau^+\tau^-)} .
\label{eq:ratioBR}
\end{eqnarray}
Such a ratio of Higgs boson decay widths is independent of the production process 
 and therefore is insensitive to higher-order QCD corrections and also to 
 new physics effects that may affect the production rate of the Higgs. 
This ratio has also the particularity of being less sensitive to 
 the systematic errors (which drop out in the ratio) 
and  could be used to discriminate the SM against other 
models such as 2HDM or supersymmetric models.

The paper is organized as follows. In Section 2 we 
review the Yukawa textures, scalar potential and Higgs self-couplings 
of the 2HDM model, as well as the theoretical and
experimental constraints on the model. 
Section 3 outlines the calculation and specifies 
the renormalization scheme we will be using. 
The numerical results are presented in Section 6. 
Finally, we conclude in Section 6.
\section{The 2HDM model}
\subsection{Yukawa textures}
In the 2HDM, fermion and gauge boson masses are generated from two Higgs
doublets $\Phi_{1,2}$ where both of them acquire vacuum expectation values
$v_{1,2}$. If both Higgs fields couple to all fermions, 
Flavor Changing Neutral Currents (FCNC) are generated
which can invalidate some low energy 
observables in  B, D and K physics. 
In order to avoid such FCNC, the Paschos-Glashow-Weinberg theorem 
\cite{weinberg} proposes a $Z_2$ symmetry that forbids FCNC couplings at
the tree level. Depending on the $Z_2$ assignment, we have 
four type of models \cite{Branco,Gunion:1989we}. 
In the 2HDM type-I model, only the second doublet $\Phi_2$ interacts with all 
the fermions like in SM. In 2HDM type-II model the doublet $\Phi_2$ interacts 
with up-type quarks and $\Phi_1$ interacts with the down-type quarks and 
charged leptons. In 2HDM type-III, 
charged leptons couple to $\Phi_1$ while all the quarks couple to $\Phi_2$. 
Finally, in 2HDM type IV, charged leptons and up-type quarks 
couple to $\Phi_2$ while down-type quarks acquire masses from their 
couplings to $\Phi_1$. 

The most general Yukawa interactions can be written as follows,
\beqa
-{\mathcal L}_\text{Yukawa}^\text{2HDM} =
&{\overline Q}_LY_u\widetilde{\Phi}_2 u_R^{}
+{\overline Q}_LY_d\Phi_dd_R^{}
+{\overline L}_LY_\ell\Phi_\ell \ell_R^{}+\text{h.c},
\label{eq:yukawah}
\eeqa
where $\Phi_{d,l}$ ($d,l=1,2$) represents $\Phi_1$ or $\Phi_2$ and 
$Y_f$ ($f=u,d$ or $\ell$) stands for Yukawa matrices.
The  two complex scalar $SU(2)$ doublets can be decomposed according to
\beqa
\Phi_i = \left( \begin{array}{c} \phi_i^+ \\
\left(v_i +  \rho_i + i \eta_i \right) \left/ \sqrt{2} \right.
\end{array} \right),
\quad i= 1, 2,
\label{eq:doubletcomponents}
\eeqa
where $v_{1,2}$ are the vacuum expectation values of $\Phi_{1,2}$.
The mass eigenstates for the Higgs bosons are obtained by 
orthogonal transformations,
\beqa
\left(\begin{array}{c}
\phi_1^\pm\\
\phi_2^\pm
\end{array}\right)&=R_\beta
\left(\begin{array}{c}
G^\pm\\
H^\pm
\end{array}\right),\quad 
\left(\begin{array}{c}
\rho_1\\
\rho_2
\end{array}\right)=R_\alpha
\left(\begin{array}{c}
H^0\\
h^0
\end{array}\right) ,\quad 
\left(\begin{array}{c}
\eta_1\\
\eta_2
\end{array}\right)
=R_\beta\left(\begin{array}{c}
G^0\\
A^0
\end{array}\right) ,
\eeqa
with the generic orthogonal matrix 
$$R_\theta = 
\left(
\begin{array}{cc}
\cos\theta & -\sin\theta\\
\sin\theta & \cos\theta
\end{array}\right). $$
\begin{table}
 \begin{center}
  \begin{tabular}{||l|l|l|l|l|l|l|l|l|l||}
   \hline \hline
type    & $\xi_u^{h^0}$ & $\xi_d^{h^0}$ & $\xi_l^{h^0}$ & $\xi_u^{H^0}$ & $\xi_d^{H^0}$ & $\xi_l^{H^0}$ & $\xi_u^{A^0}$ & $\xi_d^{A^0}$ & $\xi_l^{A^0}$ \\ \hline
    I & $c_\alpha/s_\beta$ & $c_\alpha/s_\beta$& $c_\alpha/s_\beta$ & $s_\alpha/s_\beta$ & $s_\alpha/s_\beta$ & $s_\alpha/s_\beta$ & $\cot\beta$ & 
    $-\cot\beta$ & $-\cot\beta$ \\ \hline
    II & $c_\alpha/s_\beta$ & $-s_\alpha/c_\beta$& $-s_\alpha/c_\beta$ & $s_\alpha/s_\beta$ & $c_\alpha/c_\beta$ & $c_\alpha/c_\beta$ & $\cot\beta$ & 
    $\tan\beta$ & $\tan\beta$ \\ \hline 
III & $c_\alpha/s_\beta$ & $c_\alpha/s_\beta$& $-s_\alpha/c_\beta$ & $s_\alpha/s_\beta$ & $s_\alpha/s_\beta$ & $c_\alpha/c_\beta$ & $\cot\beta$ & 
    $-\cot\beta$ & $\tan\beta$ \\ \hline
 IV & $c_\alpha/s_\beta$ & $-s_\alpha/c_\beta$& $c_\alpha/s_\beta$ & $s_\alpha/s_\beta$ & $c_\alpha/c_\beta$ & $s_\alpha/s_\beta$ & $\cot\beta$ & 
    $\tan\beta$ & $-\cot\beta$ \\ \hline \hline
    \end{tabular}
 \end{center}
\caption{Yukawa coupling coefficients of the neutral Higgs bosons $h^0,H^0,A^0$ 
to the up-quarks, down-quarks and the charged leptons ($u,d,\ell$) 
in the four 2HDM types.}
\label{Ycoupl}
\end{table}

From the eight fields initially present in the two scalar doublets, 
three of them, namely the Goldstone bosons 
$G^\pm$ and $G^0$, are eaten
 by the longitudinal components of $W^\pm$ and $Z$, respectively. 
The remaining five are physical Higgs fields,
two CP-even $H^0$ and $h^0$, a CP-odd $A^0$, and a pair of charged scalars $H^\pm$.

Writing the Yukawa interactions eq.~(\ref{eq:yukawah}) in terms of 
 mass eigenstates of the neutral and charged Higgs bosons yields 
\beqa
-{\mathcal L}_\text{Yukawa}^\text{2HDM} &=&
 \sum_{f=u,d,\ell} \frac{m_f}{v} \left(
\xi_f^{h^0} {\overline f}f h^0
+ \xi_f^{H^0}{\overline f}f H^0
- i \xi_f^{A^0} {\overline f}\gamma_5f A^0
\right)    \\ & &
+\left\{\frac{\sqrt2V_{ud}}{v}\,
\overline{u} \left( m_u \xi_u^{A^0} \text{P}_L
+ m_d \xi_d^{A^0} \text{P}_R \right)
d H^+
+\frac{\sqrt2m_\ell\xi_\ell^{A^0}}{v}\,
\overline{\nu_L^{}}\ell_R^{}H^+
+\text{h.c}\right\},    \nonumber
\label{Eq:Yukawa}
\eeqa
where $v^2=v_1^2+v_2^2=(\sqrt{2} G_F)^{-1}$; 
$P_{R}$ and $P_{L}$ are the 
right- and left-handed  projection operators, respectively. 
The coefficients $\xi^{h^0}_f, \xi^{H^0}_f$ and  $\xi^{A^0}_f$ 
($f=u, d, l$) in the four 
2HDM types are given in the Table~\ref{Ycoupl}.\\

\subsection{Scalar potential and self-coupling of the Higgs bosons}
The most general 2HDM scalar potential which is invariant under
$SU(2)_L\otimes U(1)_Y$ and possesses a soft $Z_2$ 
breaking term ($m_{12}^2$) \cite{Gunion:1989we,Branco,Gunion:2002zf} 
can be written in the following way, 
\begin{eqnarray}
  {V}_{\rm 2HDM}  &=&  m_{11}^2  \Phi_1^{\dagger}  \Phi_1
                          + m_{22}^2  \Phi_2^{\dagger}  \Phi_2  - 
                              m_{12}^2 \left( \Phi_1^{\dagger} \Phi_2 
                                + \Phi_2^{\dagger} \Phi_1 \right) 
                                \nonumber  
                       + \frac{\lambda_1}{2} 
                               (\Phi_1^{\dagger}  \Phi_1 )^2 
                             + \frac{\lambda_2}{2} 
                               (\Phi_2^{\dagger}  \Phi_2 )^2 \nonumber \\
& & + \lambda_3 (\Phi_1^{\dagger}  \Phi_1) (\Phi_2^{\dagger}  \Phi_2)
                      + \lambda_4 
                               \left| \Phi_1^{\dagger} \Phi_2 \right|^2
                             + \frac{\lambda_5}{2} 
                             \left\{ 
                               \left( \Phi_1^{\dagger} \Phi_2 \right)^2
                            +  \left( \Phi_2^{\dagger} \Phi_1 \right)^2
                             \right\} .
\label{higgspot}                             
\end{eqnarray}
Hermiticity of the potential requires 
$m_{11}^2$, $m_{22}^2$ and $\lambda_{1,2,3,4}$  to be real, 
while $m_{12}^2$ and $\lambda_5$ could be complex in case one would allow for
CP violation in the Higgs sector. In what follows we assume that there is no CP
violation, which means $m_{12}^2$ and $\lambda_5$ are taken as real.

From the above potential, Eq.~(\ref{higgspot}), we can derive the 
triple Higgs couplings, needed for the present study
as a function of the 2HDM parameters
$m_{h^0}$, $m_{H^0}$, $m_{A^0}$, $m_{H^\pm}$, $\tan\beta$, $\alpha$
and $m_{12}^2$. 
These couplings follow from the scalar potential and are thus
independent of the Yukawa types used; they are given by
\begin{eqnarray}
\lambda_{h^0h^0h^0}^{2HDM} &=& \frac{-3g}{2m_W s^2_{2\be}}\bigg[(2c_{\al+\be} + s_{2\al}
  s_{\be-\al})s_{2\be} m^2_{h^0} - 4c^2_{\be-\al} c_{\be + \al} m^2_{12}\bigg]
\label{lll}\nonumber\\
\lambda_{H^0h^0h^0}^{2HDM} &=& -\frac{1}{2}\frac{g c_{\be-\al}}{m_W s^2_{2\be}}\bigg[
  (2 m^2_{h} + m^2_{H^0}) s_{2\al} s_{2\be} -2 (3 s_{2\al}-s_{2\be})
  m^2_{12}\bigg] \nonumber \\
\lambda_{h^0H^0H^0}^{2HDM} &=& \frac{1}{2}\frac{g s_{\be-\al}}{m_W s^2_{2\be}}\bigg[
  (m^2_{h^0} + 2 m^2_{H^0}) s_{2\al} s_{2\be} - 2 (3 s_{2\al}+s_{2\be})
  m^2_{12}\bigg] \label{hhl}\nonumber\\
 \lambda_{ h^0H^\pm H^\mp}^{2HDM} &=& \frac{1}{2}\frac{g}{m_W}\bigg[
  (m^2_{h^0} - 2 m^2_{H^\pm})s_{\be-\al} - 
\frac{2c_{\be + \al}}{s^2_{2\be}}(m^2_{h} s_{2\be}-
 2 m^2_{12})\bigg] \label{hhp}\nonumber\\
 \lambda_{h^0A^0A^0}^{2HDM}  &=& \frac{1}{2}\frac{g}{m_W}\bigg[
  (m^2_{h} -2 m^2_{A^0} )s_{\be-\al} - \frac{2c_{\be + \al}}{s^2_{2\be}}(m^2_{h} s_{2\be}-
  2 m^2_{12})\bigg], 
\end{eqnarray} 
with the $W$ boson mass $m_W$ and the $SU(2)$ gauge coupling constant $g$.
We have used the notation $s_x$ and $c_x$ as short-hand notations for 
$\sin(x)$ and $\cos(x)$, respectively. 
The mixing angle $\beta$ is defined by via $\tan\beta=v_2/v_1$, where 
$v_i$ are the vacuum expectation values of the Higgs fields~$\Phi_i$.

It has been shown that the 2HDM has a decoupling limit which is reached 
for $\cos(\beta-\alpha)=0$ and $m_{H^0,A^0,H^\pm}\gg m_Z$
\cite{Gunion:2002zf}. In this limit, the coupling of the CP-even $h^0$ to SM
particles completely mimic the SM Higgs couplings including the 
 triple coupling $h^0h^0h^0$. 
Moreover, the model possesses also an alignment
limit \cite{Carena:2013ooa}, in which one of the CP-even Higgs bosons 
$h^0$ or $H^0$ looks like SM Higgs particle if $\sin(\beta-\alpha)\to 1$ or
$\cos(\beta-\alpha)\to 1$. \\
In the limit  $\alpha = \beta-\pi/2$
(which will be used for our numerical analysis) the above triple Higgs couplings
reduce to the simplified form

\begin{eqnarray} 
  \lambda_{h^0h^0h^0}^{2HDM} &=& \frac{-3g}{2m_W}m^2_{h^0} = \lambda_{hhh}^{SM},
  \label{lll} \nonumber \\
  \lambda_{H^0h^0h^0}^{2HDM} &=&0 , \nonumber \\
  \lambda_{h^0H^0H^0}^{2HDM} &=& \frac{g}{m_W}\bigg[
  \bigg(\frac{2m^2_{12}}{s_{2\be}}-m^2_{H^0}\bigg) - \frac{m^2_{h^0}}{2}
  \bigg], \label{hhl} \nonumber \\
  \lambda_{h^0H^\pm H^\mp}^{2HDM} &=&  \frac{g}{m_W}\bigg[
  \bigg(\frac{2m^2_{12}}{s_{2\be}}-m^2_{H^\pm}\bigg)-\frac{m^2_{h^0}}{2}\bigg], 
  \label{hhp}\nonumber\\
 \lambda_{h^0A^0A^0}^{2HDM}  &=& \frac{g}{m_W}\bigg[
  \bigg(\frac{2m^2_{12}}{s_{2\be}}-m^2_{A^0}\bigg)-\frac{m^2_{h^0}}{2}\bigg] ,
\end{eqnarray} 
where we can see that in the degenerate case, $m_{H^\pm}=m_{H^0}=m_{A^0}=m_{S}$,
all triple Higgs couplings $h^0H^0H^0$, $h^0A^0A^0$ and 
$h^0H^\pm H^\mp$ have the same expression, labeled by $h^0SS$,
\begin{eqnarray} 
 \lambda_{h^0SS}^{2HDM}  &=& \frac{g}{m_W}\bigg[
  \bigg(\frac{2m^2_{12}}{s_{2\be}}-m^2_{S^0}\bigg)-\frac{m^2_{h^0}}{2}\bigg].
\label{eq:hSS}
\end{eqnarray}
 
\subsection{Theoretical and experimental constraints}
The 2HDM has several theoretical constraints 
which we briefly address here. In order 
 to ensure vacuum stability of the 2HDM, the scalar potential 
must satisfy conditions that guarantee that its bounded from below,
i.e.\ that the requirement ${V}_{\rm 2HDM}\geq 0$ is satisfied for all directions
of $\Phi_1$ and $\Phi_2$ components. This requirement imposes
the following  conditions on the coefficients $\lambda_i$ \cite{BFB1,BFB2}:
\beqa
\lambda_1>0 \quad , \quad \lambda_2>0 \quad , \quad  \lambda_3 + 2
 \sqrt{\lambda_1 \lambda_2} > 0 \quad \quad , \quad \quad
 \lambda_3 + \lambda_4 - |\lambda_5| > 2 \sqrt{\lambda_1 \lambda_2}. 
\eeqa
In addition to the constraints from positivity of the scalar potential, 
there is another set of constraints by requiring perturbative
tree-level unitarity for scattering of  Higgs bosons and 
longitudinally polarized gauge bosons. These constraints are taken 
from~\cite{arhrib:unit,kanemura:unit}. Moreover,
we also force the potential to be perturbative by imposing that 
all quartic coefficients of the scalar potential
satisfy $|\lambda_i| \leq 8 \pi$ ($i=1,...,5$).

Besides these theoretical bounds, we have indirect experimental 
constraints from $B$~physics observables 
on 2HDM parameters such as $\tan\beta$ and the charged Higgs boson  mass.
It is well known that in the framework of 2HDM-II and IV, for example, 
the measurement of the $b\to s\gamma$ branching ratio requires the charged Higgs 
boson mass to be heavier than 580~GeV~\cite{Misiak:2017bgg,Misiak:2015xwa} 
for any value of $\tan\beta \geq 1$. 
Such a limit is much lower for the other 2HDM types~\cite{Enomoto:2015wbn}. 
In 2HDM-I and III, as long as $\tan\beta\geq 2$, 
it is possible to have charged Higgs bosons as 
light as 100 GeV \cite{Enomoto:2015wbn,Arhrib:2016wpw}  while being consistent 
with all $B$ physics constraints as well as with LEP
 and LHC limits \cite{Aad:2014kga,Khachatryan:2015qxa,Khachatryan:2015uua,
Aad:2013hla, Abbiendi:2013hk,Akeroyd:2016ymd}. \\
We stress in passing that after the Higgs-like particle discovery, 
several theoretical studies have performed global-fit analyses for the 2HDM 
to pin down the allowed regions of parameter space both for a SM-like 
Higgs $h^0$
\cite{Ferreira:2011aa} as well as for a SM-like Higgs boson $H^0$ \cite{H-like}.

\begin{table}[!ht]
\caption{Combined best-fit signal strengths $\widehat{\mu}_{\rm{1}}$ and
$\widehat{\mu}_{\rm{2}}$
and the associated correlation coefficient $\rho$ for corresponding Higgs decay
 mode~\cite{ATLAS:CONF}.}
\begin{center}
\begin{tabular}{|c|cccc|}
\hline\hline
$f$ & $\widehat{\mu}^{f}_{\rm{1}}$ & $\widehat{\mu}^{f}_{\rm{2}}$ 
&$\pm\,\,1\widehat{\sigma}_{\rm{1}}$
&  $\pm\,\,1\widehat{\sigma}_{\rm{2}}$ \\ \hline
$\gamma\gamma$ & $1.16 $ & 0.18 & 0.16 & 0.7 \\ \hline
$ZZ^*$ & $1.70 $ & 0.3 & 0.4 & 1.20 \\ \hline
$WW^*$ & $0.98 $ & 1.28 & 0.28 & 0.55 \\ \hline
$\tau^+\tau^-$ & $2 $ & 1.24 & 1.50 & 0.59 \\ \hline
$b\bar{b}$ & 1.11 & 0.92 & 0.65 & 0.38 \\\hline\hline
\end{tabular}
\label{tab:Higgs data}
\end{center}
\end{table}

Moreover, we take into account 
experimental data from the observed cross section times branching ratio 
divided by SM predictions for the various channels,
i.e.\ the signal strengths of the Higgs boson defined by
\begin{eqnarray}
\mu^f_{i}&=&\frac{\sigma(i \rightarrow h^0)^{2HDM} Br(h^0\rightarrow
  f)^{2HDM}}{\sigma(i\rightarrow h^0)^{SM} Br(h^0\rightarrow f)^{SM}},
\quad\quad\quad i=1,2
\end{eqnarray}
where $\sigma(i \rightarrow h^0)$ denotes the Higgs-boson production cross 
section through channel $i$ and  $Br(h^0\rightarrow f)$  is the branching
ratio  for the Higgs
decay $h^0\rightarrow f\bar{f}$. Since several Higgs production channels are 
available at the LHC, 
they are grouped to be 
$\mu^f_{1} = \mu^f_{ggF+tth^0}$ and  $\mu^f_{2} = \mu^f_{VBF+Vh^0}$,
containing gluon fusion (ggF) plus associated Higgs production $t\bar{t}h^0$, 
and vector boson fusion (VBF) plus 
Higgs-strahlung $Vh^0$ with $V= Z/W$. 
We summarize relevant signal strengths associated to each Higgs production and decay channels in Table~\ref{tab:Higgs data} with the overall combinations obtained by the ATLAS and CMS collaborations.

\section{One-loop calculation and renormalization scheme}
Calculations of higher order corrections in perturbation theory in general
lead to ultra-violet (UV) divergences.
The standard procedure to eliminate these UV divergences consists in
renormalization of the bare Lagrangian by redefinition of couplings and
fields. In the SM, the on-shell  renormalization scheme is 
well elaborated~\cite{Hollik:b,Hollik,Denner:1991kt}. For the 2HDM, 
several extensions of the SM renormalization scheme exist in the 
literature \cite{Kanemura:2014dja,Kanemura:2004mg, 
Arhrib:2003ph,Hessenberger:2016atw}. Recently, 
gauge independent renormalization schemes have been proposed 
\cite{Denner:2016etu,Krause:2016oke}, with e.g.\
$\overline{\rm{MS}}$ renormalization for the
mixing angles and the soft $Z_2$ breaking term in 
the Higgs sector~\cite{Denner:2016etu}.
In the present study, we adopt  the on-shell 
renormalization scheme used also in  \cite{Arhrib:2003ph},
which is an extension of the on-shell scheme
of the SM: the gauge sector is renormalized in analogy to 
\cite{Hollik:b,Hollik} concerning vector-boson masses and 
field renormalization; also fermion mass and field renormalization
is treated in an analogous way (see also~\cite{Denner:1991kt}).
For renormalization of the Higgs sector we take over the approach 
used in~\cite{Arhrib:2003ph}, which means
on-shell renormalization 
\begin{itemize}
\item
for the $h^0, H^0$ tadpoles, yielding zero for the renormalized tadpoles 
and thus $v_{1,2}$ at the minimum of the potential also at one-loop order,
\item
for all physical masses from the Higgs potential, defining the masses
$m_{h^0}, m_{H^0}, m_{A^0}, m_{H^\pm}$ as pole masses, 
\end{itemize}
whereas Higgs field renormalization is done in the ${\overline{\rm MS}}$ scheme. 
We assign renormalization constants $Z_{\Phi_i}$ 
for the two Higgs doublets in~({\ref{eq:doubletcomponents})  
and counter-terms for the $v_i$ 
within the doublets, according to
\begin{eqnarray}
& &\Phi_i \rightarrow (Z_{\Phi_i})^{1/2}  \Phi_i \qquad , \qquad 
v_i \rightarrow  v_i-\delta v_i \, ,
\label{reno}
\end{eqnarray}
and expand the $Z$ factors
$Z_{\Phi_i}=1+\delta Z_{\Phi_i}$ to one-loop order. 
The  $\overline{\rm{MS}}$ condition yields the field renormalization
constants as follows for all types of models listed in Table~\ref{Ycoupl}:
\beqa
\label{Zfactors}
\delta Z_{\Phi_1}^{\overline{\rm MS}} & = &
\frac{\Delta}{32\pi^2} \biggl\{
  \frac{-g^2}{m_W^2t_{\beta}}
\biggl(\xi_l^{A^0}\biggl[m_e^2 + m_\mu^2 + m_\tau^2\biggr](1+t_{\beta}\xi_l^A) + N_C\xi_d^{A^0}\biggl[m_b^2 + m_d^2 + m_s^2\biggr](1+t_{\beta}\xi_d^{A^0})  \nonumber \\
& & -N_C\xi_u^{A^0}\biggl[m_c^2 + m_t^2 + m_u^2\biggr](1-t_{\beta}\xi_u^{A^0})\biggr)+(3 g^2 + g^{\prime 2}) \biggr \} \, , \nonumber \\
\delta Z_{\Phi_2}^{\overline{\rm MS}} & = &
\frac{\Delta}{32\pi^2} \biggl \{
 \frac{-g^2}{m_W^2}\biggl(-\xi_l^{A^0}\biggl[m_e^2 + m_\mu^2 + m_\tau^2\biggr](t_{\beta}-\xi_l^{A^0}) -N_C\xi_d^{A^0}\biggl[m_b^2 + m_d^2 + m_s^2\biggr](t_{\beta}-\xi_d^{A^0})  \nonumber \\
& & +N_C\xi_u^{A^0}\biggl[m_c^2 + m_t^2 + m_u^2\biggr](t_{\beta}+\xi_u^{A^0})\biggr)  
+(3 g^2 + g^{\prime 2})\biggr\}  \, ,
\eeqa
with $\Delta =2/(4-D)-\gamma+{\rm log} 4\pi $ from dimensional
regularization, the color factor $N_C=3$ for quarks and 
$N_C=1$ for leptons,  and the gauge couplings $g$ and $g'$.
The factors $\xi_{u,d,l}^{A^0}$ can be found in
Table~\ref{Ycoupl}.
Eq.~(\ref{Zfactors}) is a generalization of the work of~\cite{Arhrib:2003ph}
with respect to the various Yukawa structures of the 2HDM.  

The renormalized self-energy of the SM-like Higgs field $h^0$ 
is the following finite combination of the unrenormalized self-energy
and counter-terms,
\begin{eqnarray}
& & \widehat{\Sigma}_{h^0}(k^2)={\Sigma}_{h^0}(k^2)-
\delta m^2_{h^0}+(k^2- m^2_{h^0})\, \delta Z_{h^0} 
\end{eqnarray}
with the on-shell mass counter-term $\delta m^2_{h^0}$
and 
$\delta Z_{h^0}=s_{\alpha}^2 \delta Z^{\overline{\rm MS}}_{\Phi_1}
   +c_{\alpha}^2 \delta Z^{\overline{\rm MS}}_{\Phi_2} $.

Owing to the $\overline{\rm{MS}}$  field renormalization, 
a finite wave function renormalization has to be assigned 
to each external $h^0$ in a physical amplitude.
This quantity is determined by
the derivative of the renormalized self-energy $\widehat{\Sigma}_{h^0}^{\prime}$ 
on the mass shell, 
given by
\beqa
\widehat{\Sigma}_{h^0}^{\prime}(m_{h^0}^2)= 
\Sigma_{h^0}^{\prime}(m_{h^0}^2) + 
(s_{\alpha}^2 \delta Z_{\Phi_1}^{\overline{\rm MS}} +
c_{\alpha}^2 \delta Z_{\Phi_2}^{\overline{\rm MS}}  ). \,
\label{eq:hderivative}
\eeqa

\noindent
Application to the one-loop calculation 
for the fermionic Higgs boson decay $h^0 \rightarrow f \bar{f}$
yields the decay amplitude which can be written as follows,

\beqa
\label{amplitude}
{\cal M}_1= - 
\frac{ig m_f}{2 m_W} \sqrt{\widehat{Z}_{h^0}}\left[\xi_f^{h^0}(1+\Delta {\cal M}_1)  
+\xi_f^{H^{0}} \Delta {\cal M}_{12} \right]
\eeqa
where
\beqa
\label{oneloopterms}
\Delta {\cal M}_1 &= & V_1^{h^0 f\bar{f}} + \delta (h^0 f\bar{f}) ,\\
\Delta {\cal M}_{12} &= &\frac{\Sigma_{h^0H^0}(m_{h^0}^2)}
                              {m_{h^0}^2 - m_{H^0}^2} 
                         -\delta\alpha \, ,  \\
 \widehat{Z}_{h^0} &=& \left[ 1+\widehat{\Sigma}_{h^0}^{\prime}(m_{h^0}^2) 
            \right] ^{-1} \, .
\eeqa
$\Delta {\cal M}_1$ is the sum of the one-loop vertex diagrams 
$V_1^{h^0 f\bar{f}}$ and the vertex counter-term $\delta (h^0 f\bar{f})$,
$\Sigma_{h^0H^0}$ is the $h^0$--$H^0$ mixing, $\delta\alpha$ represents 
the counter-term for the mixing angle $\alpha$,
and $\hat{Z}_{h^0}$  is the finite wave function renormalization of the 
external $h^0$ fixed by the derivative
of the renormalized self-energy specified above in~(\ref{eq:hderivative}).
Given the fact that the mixing angle
$\alpha$ is an independent parameter, it can 
be renormalized in a way independent of all the 
other renormalization  conditions. 
A simple renormalization condition for $\alpha$ is to require that
$\delta \alpha$ absorbs the transition $h^0$-$H^0$
 in the non-diagonal part $\Delta {\cal M}_{12}$
of the fermionic Higgs decay amplitude. Therefore, the angle $\alpha$
is hence the CP-even Higgs-boson mixing angle also at the one-loop level, 
and the decay amplitude ${\cal M}_1$ simplifies to the 
$\Delta {\cal M}_1$ term only.

The amplitude (\ref{amplitude}) together with its ingredients
is a generalization of the work in ~\cite{Arhrib:2003ph},
extended to all charged fermions and for the various 2HDM types. 
$\Delta {\cal M}_1$ contains besides the genuine vertex corrections
the counter-term $\delta(h^0 f\bar{f})$ for Higgs-fermion-fermion
vertex, which reads as follows,
\beqa
\label{vertexCT}
\delta(h^0 f\bar{f}) = \frac{\delta m_f}{m_f} + \delta Z_V^f + 
\frac{\delta v}{v} \label{dhbb} \, ,
\eeqa
where
\beqa
\frac{\delta m_f}{m_f} + \delta Z_V^f = \Sigma_S^f(m_f^2) - 2 m_f^2 
\left[ \Sigma_S^{\prime b}(m_f^2) + \Sigma_V^{\prime f}(m_f^2) \right]
\eeqa
can be expressed 
in terms of the scalar functions of the fermion self-energy,
\beqa 
\Sigma^f(p) &=& \not{\!p}\,\Sigma_V^f (p^2) \,+\, 
                \not{\!p} \gamma_5\,\Sigma_A^f (p^2)\, + \,
                m_f\, \Sigma_S^f (p^2) \, ,
\eeqa
and the universal part
\beqa
2 \frac{\delta v}{v}=2 \frac{\delta v_{1,2}}{v_{1,2}}& =  &
c_{\beta}^2\, \delta Z_{\Phi_1}^{\overline{\rm MS}} + s_{\beta}^2\,
\delta Z_{\Phi_2}^{\overline{\rm MS}}   \\ &+&
\Sigma_{\gamma\gamma}^{\prime}(0) +
2\frac{s_W}{c_W} \frac{\Sigma_{\gamma Z}(0)}{m_Z^2}
-\frac{c_W^2}{s_W^2} \frac{\Re\Sigma_{ZZ}(m_Z^2)}{m_Z^2} +
\frac{c_W^2-s_W^2}{s_W^2} \frac{\Re\Sigma_{WW}(m_W^2)}{m_W^2} \, .
\nonumber
\eeqa
This universality is a consequence of the renormalization condition
\beqa
 \frac{\delta v_1}{v_1}  -   \frac{\delta v_2}{v_2}  & = & 0 
\label{eq:vacren}
\eeqa
(see the discussion in \cite{Arhrib:2003ph}), which is also used in the
Minimal Supersymmetric SM (MSSM),
see e.g.~\cite{Dabelstein:1994hb,Chankowski:1992er,Frank:2006yh}.}
It is important that the singular part of the 
difference in the lhs.\ of (\ref{eq:vacren}) vanishes.
The singular part of 
$\delta v$ is  $(\delta v/v)_{\overline{\rm MS}} = 
-\frac{1}{64\pi^2} (3 g^2 + g^{\prime 2}) \Delta$, which is equal
to the expression found in the MSSM and constitutes a 
check of our calculation.

\begin{figure}[t!]\centering
\includegraphics[width=0.7\textwidth]{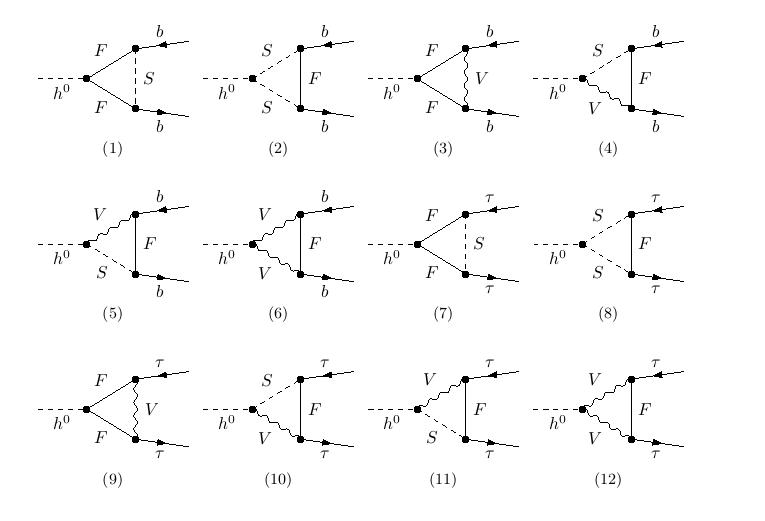}
\caption{Generic one-loop 2HDM Feynman diagrams contributing to
$\Gamma_{1}(h^0\rightarrow b \overline{b})$ and 
$\Gamma_{1}(h^0\rightarrow \tau^+ \tau^-)$.}
\label{dbt}
\end{figure}

We end this section by showing in Fig.~(\ref{dbt}) 
 the one-loop Feynman diagrams in 2HDM for 
$h^0\rightarrow b \overline{b}$ and  $h^0\rightarrow \tau^+ \tau^-$, 
where S stands for ($H^{\pm},A^0,H^0,G^{\pm})$ 
for both decays while F represents ($b$,$t$) for
 $h^0\rightarrow b \overline{b}$ and ($\tau, \nu_{\tau}$) for 
  $h^0\rightarrow \tau^+ \tau^-$. 
In the SM limit~\cite{Gunion:2002zf}, 
diagrams (4, 5, 10, 11)  and  (2, 8) with (S,S)=($H^{\pm},G^{\pm})$ vanish. 
Consequently, the important effects come from diagrams (1, 2) and (7,8) 
respectively for $h^0\rightarrow b \overline{b}$ and 
$h^0\rightarrow \tau^+ \tau^-$.

In the present work, computation of all the one-loop amplitudes and 
counter-terms is done with the help of FeynArts and FormCalc~\cite{FAFC}
packages. Numerical evaluations of the scalar integrals are done with 
LoopTools~\cite{LT}. We have  also tested the cancellation of 
UV divergences both analytically and numerically.

\section{Results}
 Before illustrating our findings, we first present the one-loop quantities 
that we are interested in. At one-loop order the decay width of 
the Higgs-boson into $b\overline{b}$ and $\tau^+ \tau^-$ is 
given by the following expressions,

\beqa
\Gamma_1(h^0 \to f\bar{f}) &=& 
\frac{N_C \alpha m_f^2}{8 s_W^2 m_W^2} \beta^3 m_{h^0} \,
( \xi_f^{h^0})^2 \,
 \widehat{Z}_{h^0} \,\biggl[1 + 2 \Re (\Delta {\cal M}_1) \biggr] \, ,
\label{width0}
\eeqa
where $\beta^2=1-4m_{f}^2/m_{h^0}^2$. We will parameterize the 
tree level width by the Fermi constant $G_F$, i.e.\ we use the relation
\beqa
\alpha&=& \frac{s_W^2 m_W^2 \sqrt{2} G_F}{\pi (1+\Delta r)}\approx
\frac{s_W^2 m_W^2 \sqrt{2} G_F}{\pi} (1-\Delta r)\
\eeqa
where $\Delta r$ incorporates higher-order corrections. 
According to the above relation, the one-loop decay width 
eq.~(\ref{width0}) becomes
\beqa
\Gamma_1(h^0 \to f\bar{f}) &=& 
\frac{N_C G_F m_f^2}{4 \sqrt{2} \pi} \beta^3 m_{h^0} \,
( \xi_f^{h^0})^2 \,
 \widehat{Z}_{h^0} \,\biggl[1-\Delta r + 2 \Re (\Delta {\cal M}_1) \biggr] \,\nonumber  \\
 &=&\Gamma_0(h^0 \to f\bar{f}) \widehat{Z}_{h^0} \,\biggl[1 - \Delta r + 
2 \Re (\Delta {\cal M}_1)\biggr]\, .
\label{width}
\eeqa
To parameterize the quantum corrections, we define the following 
one-loop ratios:
\begin{eqnarray}
 \Delta_{bb}&=&\frac{\Gamma_{1}^{2HDM}(h^0\rightarrow b 
\overline{b})}{\Gamma_{1}^{SM}(h^0\rightarrow b \overline{b})},\\
 \Delta_{\tau \tau}&=&\frac{\Gamma_{1}^{2HDM}(h^0\rightarrow 
\tau^+ \tau^-)}{\Gamma_{1}^{SM}(h^0\rightarrow \tau^+ \tau^-)} ,
\end{eqnarray}
where we also take the SM decay width 
 $\Gamma_{1}^{SM}(h\rightarrow f\bar{f})$ with the one-loop electroweak 
corrections. The two ratios defined above  will take
the following form:
\begin{eqnarray}
 \Delta_{ff}&=&\frac{ \widehat{Z}_{h^0}(1 -\Delta r^{2HDM}+ 
2 \Re (\Delta {\cal M}_1^{2HDM}) ) }{(1 -\Delta r^{SM}+ 
2 \Re (\Delta {\cal M}_1^{SM}) ) },  \ \ f=b, \tau \, .
\label{eq:deltaff}
\end{eqnarray}

Another observable that could help in distinguishing between models 
is the ratio of branching fractions as given by~\cite{Arganda:2015bta},
\begin{eqnarray}
R=BR(h^0\to b\bar{b})/BR(h^0\to \tau^+\tau^-) \, .
\end{eqnarray}
At leading order,  this ratio reads as follows,
\begin{equation}
  \label{eq:RteoSM}
  R=3\frac{m_b^2(m_{h^0})}{m_\tau^2} \times 
\left\lbrace
\begin{array}{l}
1 \ : \  \ \rm{SM} \ , \rm{2HDM\ I} \ \ and
  \ \ \rm{2HDM\ II}\\
\frac{1}{\tan^2\beta \tan^2\alpha} \ \ : \ \ \rm{2HDM\ III} \\
\tan^2\beta \tan^2\alpha \ \ : \ \ \rm{2HDM\ IV}
\end{array}
\right.
\end{equation}
where we take the running mass of the $b$ quark at $m_h$. 
Note that in the alignment limit, the above ratio $R$ 
simplifies to $R=3 m_b^2(m_{h^0})/m_\tau^2$ for the SM and for all four 2HDM types.

The ratio $R$ does not depend on the production mechanism
of the Higgs boson and is therefore insensitive to higher-order QCD corrections
 and also to any new physics that affects the production process. In addition,
this ratio is also less sensitive to systematic errors since 
 some of them drop out in the ratio.

Let us define the ratio $R^\text{2HDM}/R^\text{SM}$ in terms of the quantity
\begin{eqnarray}
X=\frac{R^\text{2HDM}}{R^\text{SM}}=\frac{
  \Delta_{bb}}{\Delta_{\tau^+\tau^-}} \, ,
\label{eq:ratio-r}
\end{eqnarray}
where we have used the same notation as in \cite{Arganda:2015bta}.
Similar to $h^0\to b\bar{b}$ and $h^0\to \tau^+\tau^-$, 
this ratio $X$ will be also sensitive to  the 
triple Higgs couplings $h^0H^0H^0, h^0A^0A^0$ and $h^0H^\pm H^\mp$ 
as well as to the Yukawa
couplings. Therefore, this ratio is a discriminating quantity between SM, 
2HDM, MSSM and other SM extensions. \\
As explained in \cite{Arganda:2015bta}, the combination of the LHC coupling
measurements can be used to extract an experimental determination of the 
$X$ ratio defined in (\ref{eq:ratio-r}),
\begin{equation}
  \label{eq:Xexp}
  X^\text{exp}=\frac{R^\text{exp}}{R^\text{SM}} =
 \frac{\lambda_{bZ}^2}{\lambda_{\tau Z}^2}, 
\end{equation}
where $\lambda_{xy}=\kappa_x / \kappa_y$.        

Both CMS and ATLAS collaborations provide~\cite{Khachatryan:2014jba}
 some values for $\lambda_{bZ}$ and $\lambda_{\tau Z}$ extracted from Higgs
 branching ratios measurements. Taking the following 
CMS and ATLAS measurements for $\lambda_{bZ}$ and $\lambda_{\tau Z}$,
\begin{equation}
\label{eq:lambdasexps}
\lambda_{bZ}^\text{CMS}=0.59_{-0.23}^{+0.22} \ \ , \ \ 
\lambda_{\tau Z}^\text{CMS}=0.79_{-0.17}^{+0.19} \ \ , \ \ 
\lambda_{bZ}^\text{ATLAS}=0.60 \pm 0.27\ \ , \ \ 
\lambda_{\tau  Z}^\text{ATLAS}=0.99_{-0.19}^{+0.23}\, ,
\end{equation}
one can get the following experimental values for $X$:
\begin{equation}
  \label{eq:XexpATLASCMS}
X^\text{CMS} = 0.56_{-0.52}^{+0.48} \ \ , \ \ 
X^\text{ATLAS} = 0.37_{-0.37}^{+0.36} \,.
\end{equation}

\begin{figure}[t!]\centering
\includegraphics[width=0.475\textwidth]{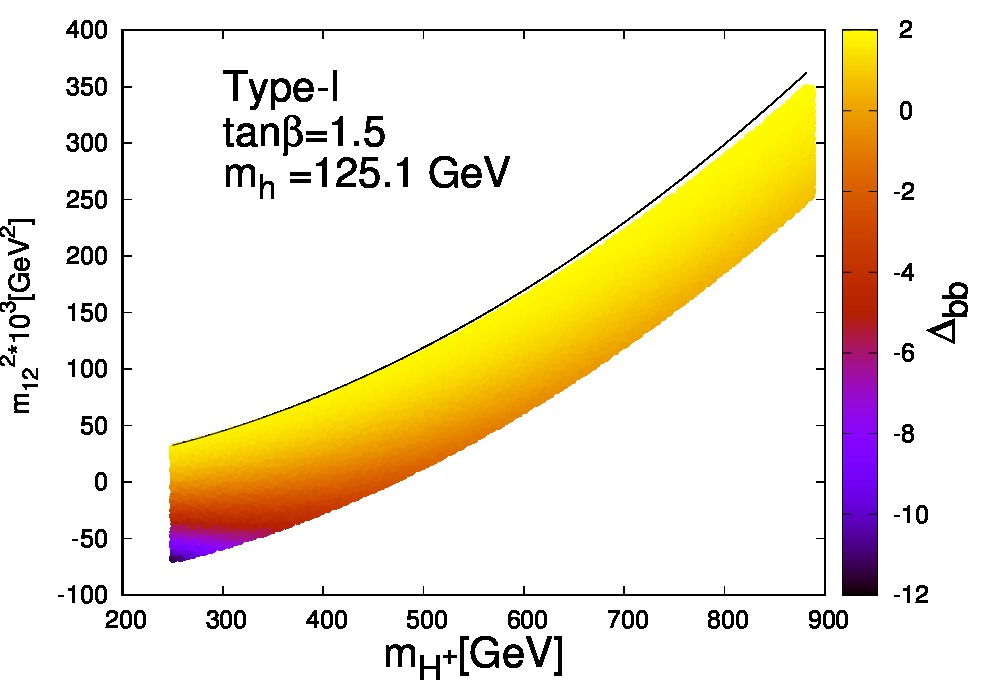}
\includegraphics[width=0.475\textwidth]{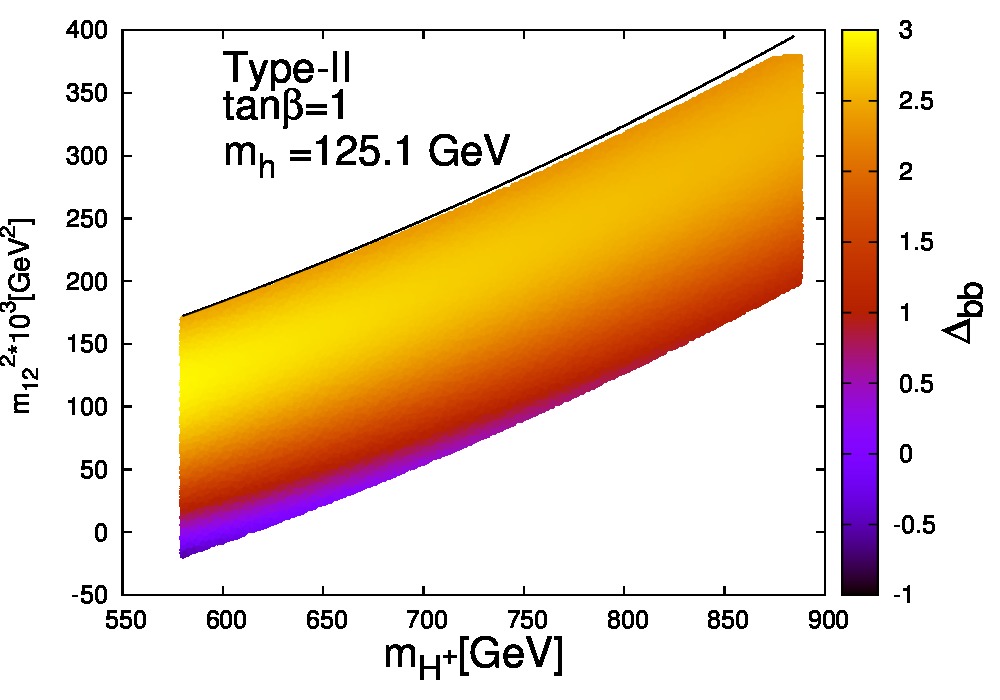}
\caption{Scatter plot for $\Delta_{bb}$ 
in $(m_{H^\pm}, m_{12}^2)$ plane for 
$\tan\beta=1.5$ in the 2HDM type I and $\tan\beta=1$ in the 2HDM  II.
 Right column shows the size of the corrections to the 
$h\to b \overline{b}$.}
\label{fig:hbb1}
\end{figure}
We have checked with \cite{Kanemura:2015mxa, Kanemura:2014dja}.
Our results slightly disagree;  presumably the small disagreement 
is due to the different renormalization schemes.
In our discussion, we will use the following SM set of parameters:
\begin{eqnarray}
&& \alpha=\frac{1}{137} \qquad , \qquad m_Z=91.1882\ \rm{GeV} 
\qquad , \qquad
m_W=80.419\ \rm{GeV} \nonumber\\
&&
 m_\tau=1.77703\  \rm{GeV} \qquad , \qquad m_b=4.7\  \rm{GeV} 
\qquad , \qquad
m_t=174.3\ \rm{GeV} \nonumber
\end{eqnarray}

For the 2HDM parameters, in order to  
simplify our analysis, we consider the alignment limit of the 2HDM, 
$\cos(\beta-\alpha)=0$, and assume that the heavy states $H^0$, $A^0$ and
$H^\pm$ are degenerate, $m_{H^{\pm}}=m_{A^0}=m_{H^0}=m_S\in [250,900]$ GeV for
2HDM type I and $m_{H^{\pm}}=m_{A^0}=m_{H^0}=m_S\in [580,900]$ GeV for 2HDM
type II. 
The CP-even $H^0$ couplings to gauge bosons $V=W,Z$
are proportional to $\cos(\beta-\alpha)$ and thus $H^0VV$ vanishes in the 
alignment limit. The CP-odd nature of $A^0$ does not allow 
$A^0$-couplings to gauge bosons. Therefore, limits from 
ATLAS and CMS \cite{Khachatryan:2015cwa}
 on heavy Higgs particles decaying to gauge bosons would be satisfied.
On the other hand, the couplings of $H^0$ and $A^0$ to a pair of $\tau$ leptons
are proportional to $\tan\beta$ and $\cot\beta$, respectively, 
in 2HDM-(II,III) and 2HDM-(I,IV). It follows that, in order not to violate
LHC data for heavy Higgs-boson decays into $\tau$ pairs, one has to keep 
 $\tan\beta$ at not too large values.\\
Moreover, in the degenerate case $m_{H^{\pm}}=m_{A^0}=m_{H^0}=m_S$, the 
electroweak precision observables are automatically satisfied, $T=0$ and 
$S=0$ ~\cite{rho} due to custodial symmetry which is preserved for 
$m_{H^{\pm}}=m_{A^0}$. It has been demonstrated recently, that at the 2 loop
level with $m_{H^{\pm}}=m_{A^0}$, the extra 2-loop contributions to $T$ still
vanish \cite{Hessenberger:2016atw}.\\
Therefore, we scan over the following range:
\begin{eqnarray}
&&m_{h^0}=125.1\ \rm{GeV}, \quad tan \beta \in [1,30], \, \, 
m_{12}^2 \in [-1\times  10^5,4\times 10^5] \ \rm{GeV}^2, \nonumber\\ 
&& m_{A^0}=m_{H^0}=m_{H^{\pm}}\in [m_{H^{\pm}}^{min},900]\ \rm{GeV},
\end{eqnarray}
$\alpha$ is fixed by the alignment limit relation $\beta-\alpha=\pi/2$.
$m_{H^{\pm}}^{min}$ is greater than 580 GeV for any value of $\tan\beta$ in 2HDM type II and IV 
\cite{Misiak:2017bgg,Misiak:2015xwa}  
while for type I and III $m_{H^{\pm}}^{min}$ could be taken 
as low as 100 GeV as long as $\tan\beta\geq 2$ \cite{Enomoto:2015wbn}. 
In our scan for 2HDM type I we take $\tan\beta\geq 1.5$ 
which constrains the  charged Higgs mass to be heavier than $250$ GeV.\\
\begin{figure}[t!]\centering
\includegraphics[width=0.5\textwidth]{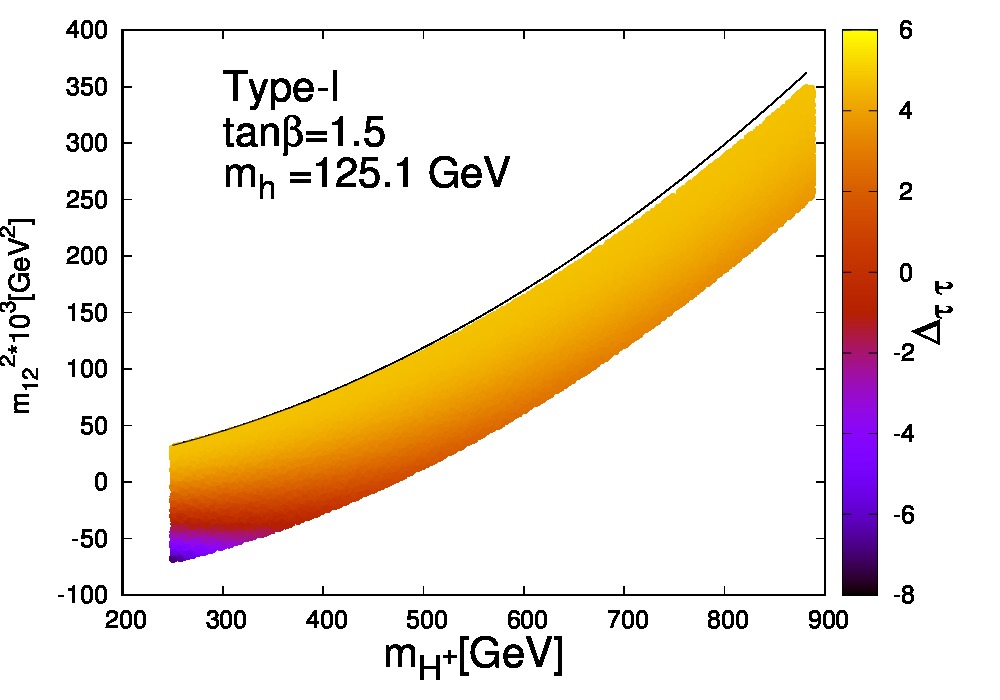}\includegraphics[width=0.5\textwidth]{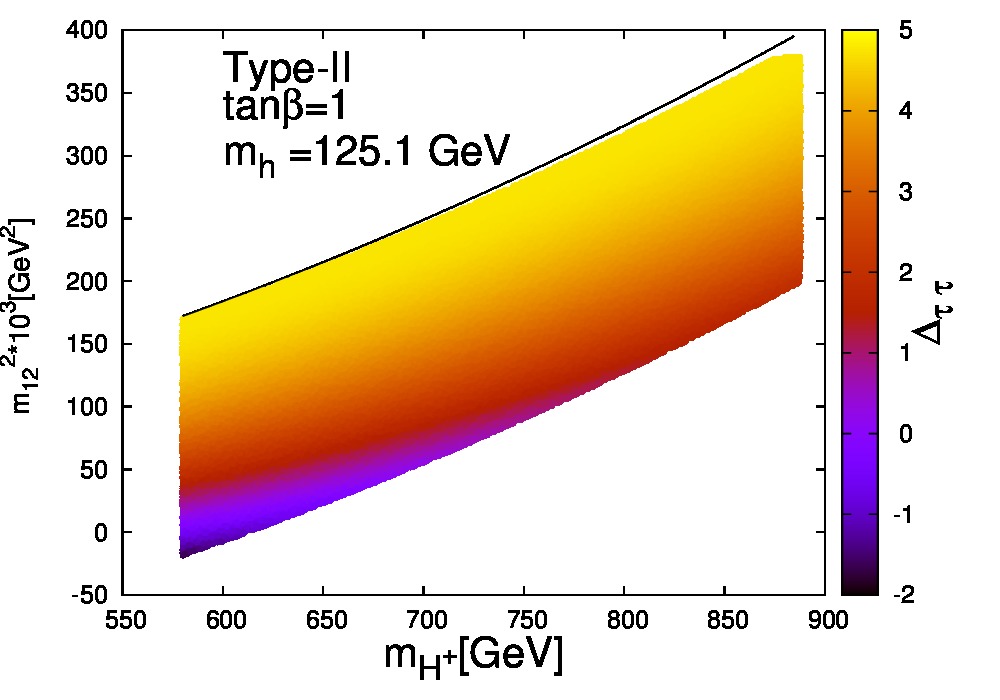}
\caption{Scatter plot for $\Delta_{\tau \tau}$ 
in $(m_{H^\pm}, m_{12}^2)$ plane for 
$\tan\beta=1.5$ in the 2HDM type I and $\tan\beta=1$ in the 2HDM  II.
Right column shows the size of the corrections to the 
$h\to \tau^+ \tau^-$.}
\label{fig:htt1}
\end{figure}

We first mention that, in the alignment limit 
with degenerate heavy Higgs particles,
the overall factor $(1-\Delta r^{2HDM})/(1-\Delta
r^{SM})$ appearing in the ratio $\Delta_{ff}$ eq.~(\ref{eq:deltaff}) 
 is close to unity since $\Delta r^{2HDM}$ and $\Delta r^{SM}$ 
becomes similar in such limit.\\
In Fig.~(\ref{fig:hbb1}) and Fig.~(\ref{fig:htt1})  
we illustrate respectively the ratios $\Delta_{bb}$ and 
$\Delta_{\tau^+\tau^-}$ in the $(m_{H^\pm}, m_{12}^2)$ plane. 
The corrections are shown in the right column in percent. 
In Fig.~(\ref{fig:hbb1}) we show only type I and
II, since in the case of $b\bar{b}$ type III and IV are 
respectively similar to type I and type II. In type II, 
these corrections are mild and could flip sign depending on
the sign of $m_{12}^2$. This means that radiative corrections effects could
either enhance $h^0\to f\bar{f}$ or suppress it with respect to SM values.
It is clear from eq.~(\ref{eq:hSS}) that 
the couplings $h^0H^0H^0$, $h^0A^0A^0$ and 
$h^0H^\pm H^\mp$  become stronger for 
negative $m_{12}^2$ where we would expect some large deviation. It is
important to notice also that the $h^0SS^{2HDM}$ ($S=A^0, H^0, H^\pm$) 
couplings would vanish for
\begin{eqnarray}
m_{12}^2&=&\frac{\sin 2\beta}{4} (m_{h^0}^2 +2 m_S^2) \quad , 
\quad S=A^0, H^0, H^\pm \, .
\label{eq:m12ms}
\end{eqnarray}
Accordingly, we expect that for such values of $m_{12}^2$ 
the loop contributions are rather small. Therefore, as a reference point, we 
display by a solid line in Fig.~(\ref{fig:hbb1}) and Fig.~(\ref{fig:htt1}) 
the parabola in
eq.~(\ref{eq:m12ms}) where the triple $h^0H^0H^0$, $h^0A^0A^0$, 
$h^0H^\pm H^\mp$ couplings vanish.

In all 2HDM types, for $m_{H^\pm}\geq 580$ GeV, 
the effects on $\Delta_{bb}$ and $\Delta_{\tau\tau}$ are rather mild in 
2HDM type (II,IV) and slightly larger in type (I,III). 
In fact, for $m_{H^0}=m_{A^0}=m_{H^\pm}\geq580$~GeV, the deviation of
$\Delta_{bb}$ is in the range $[-2\%,2\%] ([-0.5\%,3\%])$ respectively for 2HDM type-I (type-II), while in the 
case of $\Delta_{\tau\tau}$ turn out to be in the range  $[2\%,5\%] ([-1.5\%,5\%])$ respectively for 2HDM type-I (type-II).
Note that the difference between type I and II is due to
the sign change of $\xi_{d,l}^{A^0}$ couplings in type I with respect to type II.
However, for $m_{H^\pm}\leq 400$ GeV,
which is still allowed by B physics in 2HDM type I and III, 
one can see that 
$\Delta_{bb}$ and $\Delta_{\tau\tau}$ could
exceed 10\% for negative $m_{12}^2$. These
large corrections are achieved in 2HDM type I and III  
for light charged Higgs bosons as well as for negative $m_{12}^2$ where the triple Higgs
couplings $h^0SS$ ($S=A^0,H^0,H^\pm$) are enhanced. In fact, this 
enhancement is amplified with the presence of the four diagrams 
like (1)-(2) for $h^0 b{\bar b}$ and (7)-(8) for
$h^0 \tau\tau$ from Fig.~(1) with $S=H^0,A^0, H^\pm$ simultaneously 
lighter than 400 GeV. \\
On the other hand, for 2HDM type II and IV, if we still keep 
$m_{H\pm}=580$ GeV or higher in order to fulfill $b\to s\gamma$ constraint
 and relax $m_{A^0}=m_{H^0}$ to be less than 400 GeV 
therefore these light $A^0$ and $H^0$
 can induce some enhancement in $\Delta_{bb}$ and $\Delta_{\tau^+\tau^-}$
 which could reach respectively [$-12\%,6\%$] and [$-14\%,5.5\%$] for  relatively light $m_{A^0,H^0}$. 
The maximum effects is reached for 
$m_{A^0}=m_{H^0}=100$ GeV and negative $m_{12}^2$. 
The maximum effects is less than 
 in 2HDM-I and III because in the case of type-II and IV we have only 
$A^0$ and $H^0$ that could be in the range [100,200] GeV.\\

\begin{figure}[t!]\centering
\includegraphics[width=0.48\textwidth]{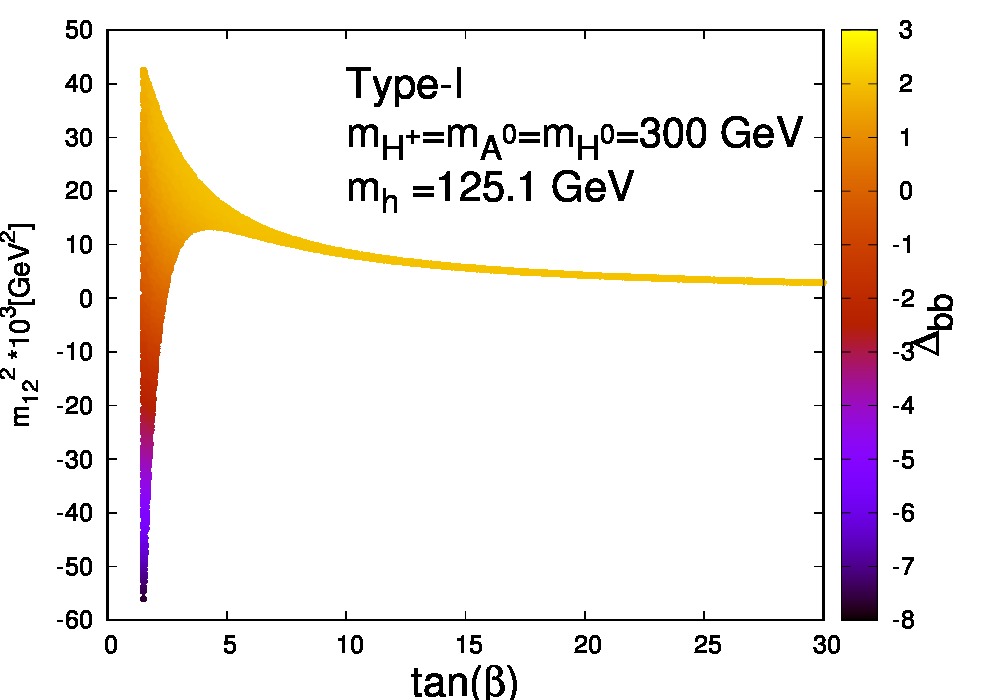}
\includegraphics[width=0.48\textwidth]{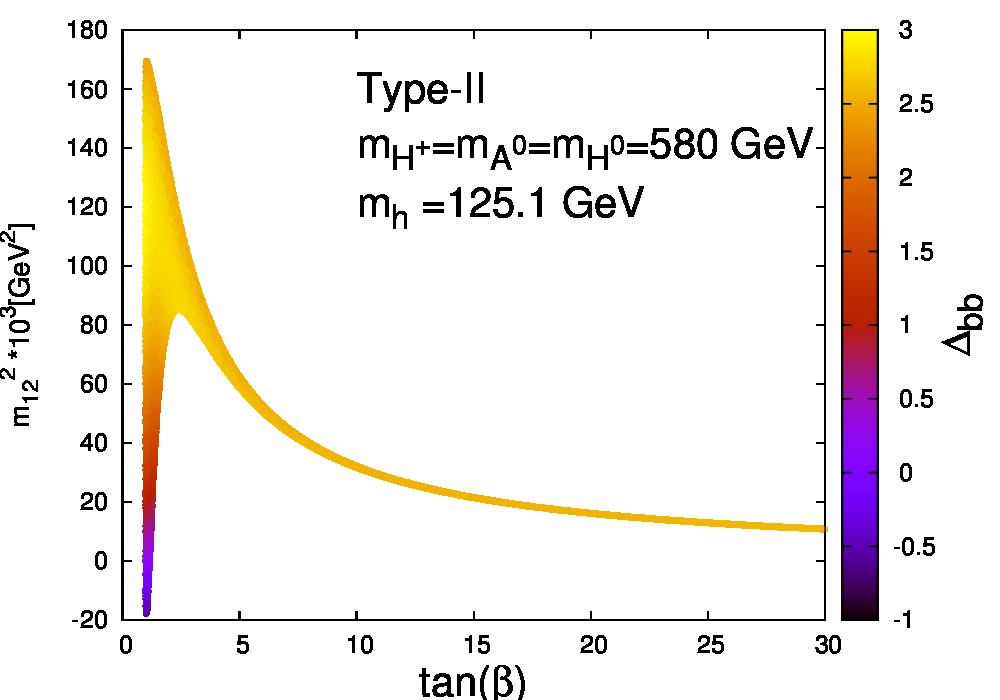}
\caption{Scatter plot for $\Delta_{bb}$ 
in $(\tan\beta, m_{12}^2)$ plane for 
$m_{H\pm}=300$ GeV in the 2HDM type I (left) and type II with $m_{H\pm}=580$ GeV (right).
Right column shows the size of the corrections to the 
$h\to b \overline{b}$.}
\label{fig:hbb2}
\end{figure}
\begin{figure}[t!]\centering
\includegraphics[width=0.48\textwidth]{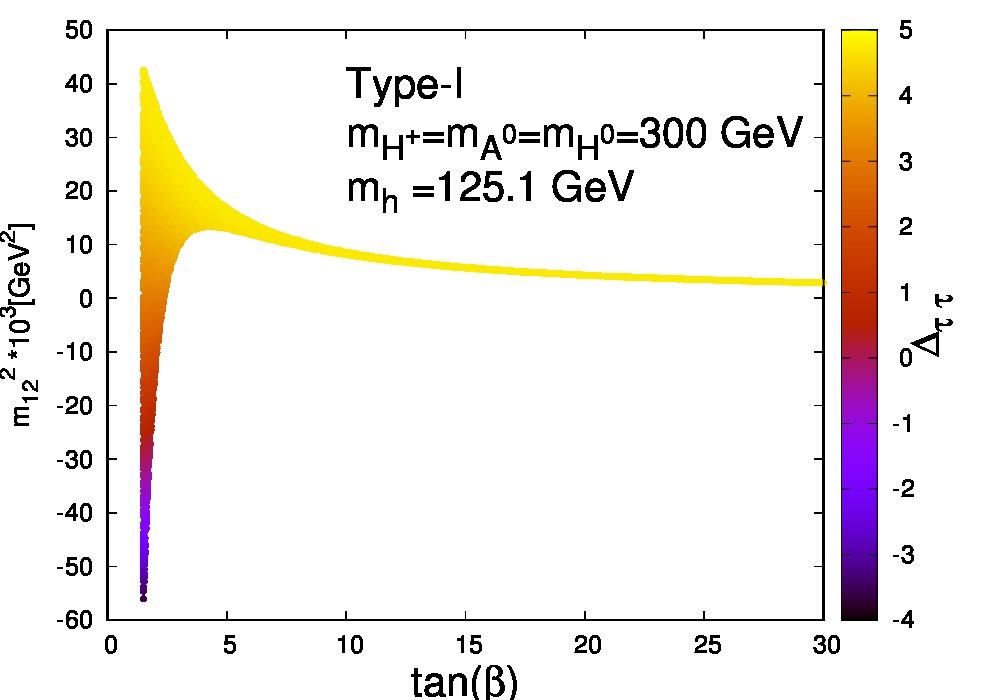}
\includegraphics[width=0.48\textwidth]{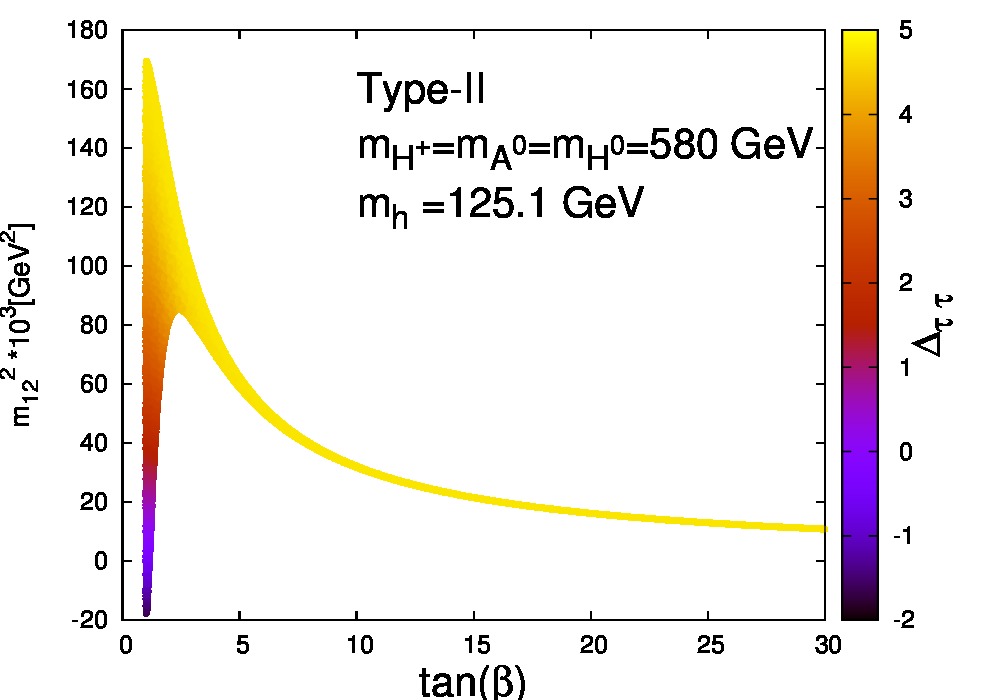}
\caption{Scatter plot for $\Delta_{\tau \tau}$ in the $(\tan\beta, m_{12}^2)$ 
plane for $m_{H^\pm}=480$ GeV in the 2HDM I (left) and
$m_{H^0}=m_{A^0}=m_{H\pm}=580$ GeV in 2HDM II (right). 
Right column shows the size of the corrections to the $h\to \tau^+ \tau^-$.}
\label{fig:htta2}
\end{figure}

In Fig.~(\ref{fig:hbb2}) and Fig.~(\ref{fig:htta2}) we show $\Delta_{bb}$ and
$\Delta_{\tau \tau}$ in the plane $(\tan\beta, m_{12}^2$) for 
 $m_{H^0}=m_{A^0}=m_{H\pm}=300$ GeV in 2HDM-I (left) and
$m_{H^0}=m_{A^0}=m_{H\pm}=580$ GeV in 2HDM-II (right).
In this scenario 
perturbative unitarity requests that $m_{12}^2$ should be small for large
$\tan\beta$. For $\tan\beta\approx 1$, the allowed range for $m_{12}^2$ 
is $[-20,170]\times 10^3$ GeV$^2$ in 2HDM type-II whilst for $\tan\beta=1.5$ the allowed range is
 $[-60,40]\times 10^3$ GeV$^2$ in the case of 2HDM type-I. For 
$\Delta_{bb}$ the corrections are between -8\% $\to$ 2\%
in 2HDM type-I and -1\% $\to$ 3\% in 2HDM type-II whereas the corrections in $\Delta_{\tau^+\tau^-}$ are in the range 
-2\%$\to$ 5\% (-4\%$\to$ 5\%) respectively for 2HDM type-I (type-II). As
explained before, this difference between type I and II is due to
the sign change of $\xi_{d,l}^{A^0}$ couplings in type I with respect to type II.\\
As we have seen previously, 
the 2HDM corrections almost decouple for 
heavy Higgs masses around 800 GeV and are of the order 3\% and 5\% 
respectively for $\Delta_{bb}$ and $\Delta_{\tau^+\tau^-}$.  
The 2\% difference between the two channels can be assigned  to the effect of  
virtual top quarks~\cite{Dabelstein:1991ky}.
In fact, in the case of $h^0\to \tau^+\tau^-$
the top effect in $\delta v/v$ and in $\Delta r$ add constructively 
while in the case of  $h^0\to b\bar{b}$ there is also a top contribution coming
from the vertex corrections which cancels part of the universal top 
contribution in $\delta v/v$ and in $\Delta r$.

\begin{figure}[t!]\centering
\includegraphics[width=0.5\textwidth]{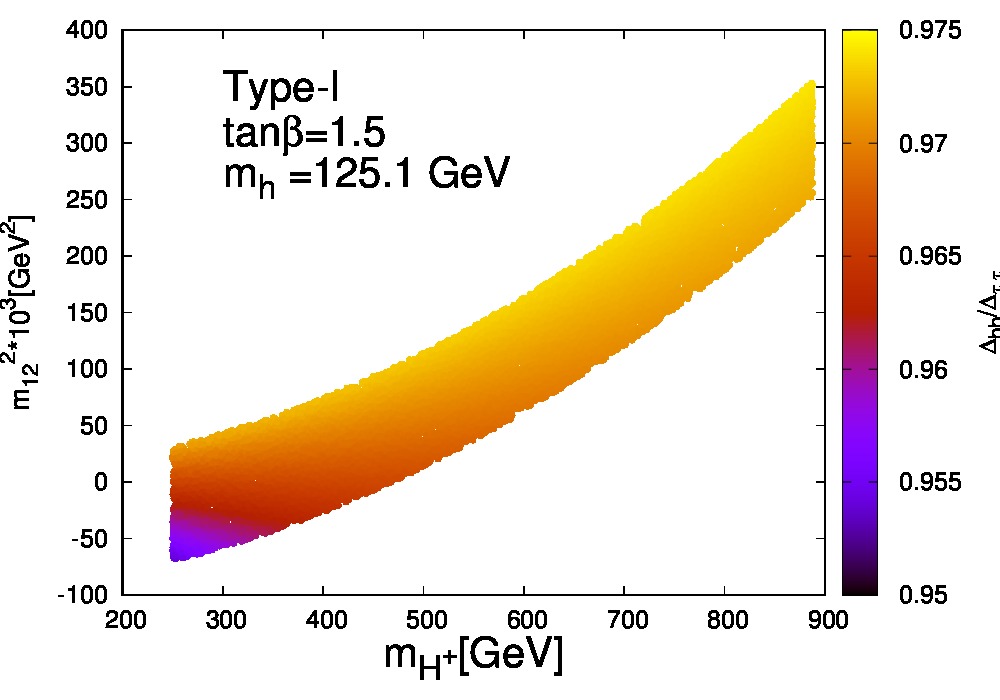}\includegraphics[width=0.5\textwidth]{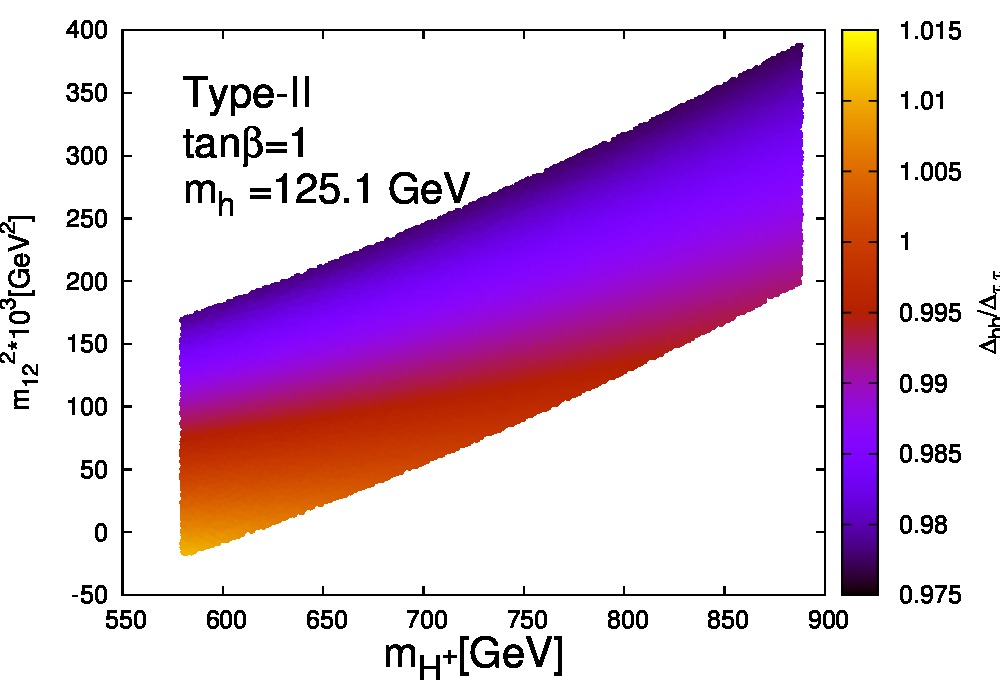}\\
\caption{Scatter plot for $X=\frac{\Delta_{bb}}{\Delta_{\tau \tau}}$ 
in the plane $(M_{H^\pm}, m_{12}^2)$ for $\tan\beta=1$ in the 2HDM-I and II}.
\label{fig:hbb3}
\end{figure}

We now proceed to discuss the effects of the triple Higgs couplings on the ratio 
$R$ defined through eqs.~(\ref{eq:ratio-r}). As explained previously, 
it is of advantage to consider the ratio-of-ratios $X$ introduced in
eq.~(\ref{eq:ratio-r}). 
The ratio $X$ is illustrated in Fig.~(\ref{fig:hbb3}) as a scatter plot 
in the plane ($m_S, m_{12}^2)$ in the alignment limit and with 
$\tan\beta=1.5$
for 2HDM type-I and $\tan\beta=1$ for 2HDM type-II. 
We obtain similar effects for 2HDM type III and IV. 
 It can be read from the plot that in the 2HDM type-II the ratio $X$
deviates from unity by about 2\% at best. This is of course a consequence
 of the fact that $h^0\to b\bar{b}$ and $h^0\to \tau^+\tau^-$ do not receive 
 significant corrections from $hSS$ in the degenerate case 
$m_{H^0}=m_{A^0}=m_{H^\pm}=m_S$.
In 2HDM type I, we have seen that $h^0SS$ modify the $h^0\to b\bar{b}$ and 
$h^0\to \tau^+ \tau^-$ decay significantly. This translates into an effect of 
 the order 5\% in the ratio X ,
which can bee seen for $m_S\approx 250$ GeV and negative $m_{12}^2$.
Notice also that in 2HDM type I, the $X$ ratio is always less than one
 while in type II it could be both, larger than one and smaller than one.\\
On the other hand, in the nondegenerate case, 
in the 2HDM II with charged Higgs-boson  mass 580 GeV and
 the neutral heavy states $m_{H^0}=m_{A^0}\in [200,400]$ GeV 
the ratio $X$ is in the range $[0.97,1]$ which does not deviate too 
much from the degenerate case.

\section{Conclusion}
We have evaluated the radiative corrections to the decays $h^0\to b\bar{b}$ and 
$h^0\to \tau^+ \tau^-$ in the framework of 2HDM type I, II, III and IV. 
Such models accommodate in their spectrum a CP-even Higgs which completely 
mimic the SM-Higgs-like seen by ATLAS and CMS
at the LHC. We have used an on-shell renormalization scheme
 for all parameters except for wave function renormalization of the Higgs
 doublet which has been done in the $\overline{\rm{MS}}$ scheme.
We performed our numerical analysis in the 
alignment limit of the 2HDM $\sin(\beta-\alpha)=1$  for 
masses $m_{H^0,A^0,H^\pm} \in [250, 800]$ GeV. 
We have shown that in type II and IV the electroweak radiative 
corrections are rather small once we take into
account  that the heavy states $A^0$, $H^0$ and $H^\pm$ have a 
mass greater than 580 GeV while it could be slightly larger for 2HDM type I
and III. We also discussed the impact of the triple Higgs couplings on the ratio 
of branching fraction $X$ and show that their effects are rather mild;
in the ratio $X$ they are smaller than in case of the MSSM~\cite{Arganda:2015bta}. \\
We conclude that at the LC, where it is expected that Higgs 
couplings to fermions can be measured with percent level precision, 
it would be possible to distinguish between various 2HDM models by 
looking at these quantum effects in Higgs observables which are shown 
here to be larger than few percent in specific cases.

\section*{Acknowledgments}
This work is supported by the Moroccan Ministry of Higher
Education and Scientific Research MESRSFC and  CNRST: Projet PPR/2015/6.
AA and RB would like to acknowledge the hospitality of the 
National Center for Theoretical Sciences (NCTS), Physics Division in Taiwan.


\bibliographystyle{unsrt}

\end{document}